\documentclass[twocolumn,aps,prd,epsf]{revtex4}
\usepackage{epsfig}
\usepackage{amssymb}
\usepackage{subfigure}

\newcommand{\be}{\begin{equation}}
\newcommand{\ee}{\end{equation}}
\newcommand{\bea}{\begin{eqnarray}}
\newcommand{\eea}{\end{eqnarray}}

\begin{document}

\title{ 
Confinement of monopole field lines in a superconductor at $T\neq 0$
}
\author{Marco Cardoso$^{1,2,3}$}
\email{mjdcc@cftp.ist.utl.pt}
\author{Pedro Bicudo$^{1,2}$} 
\email{bicudo@ist.utl.pt}
\author{Pedro D. Sacramento$^{1,3}$}
\email{pdss@cfif.ist.utl.pt}
\affiliation{ 
$^1$ Departamento de F\'{\i}sica, Instituto Superior T\'ecnico,
Av. Rovisco Pais, 1049-001 Lisboa, Portugal 
\\ 
$^2$ Centro de F\'{\i}sica Te\'orica de Part\'{\i}culas, Instituto Superior T\'ecnico,
Av. Rovisco Pais, 1049-001 Lisboa, Portugal
\\ 
$^3$ Centro de F\'{\i}sica das Interac\c{c}\~oes Fundamentais, Instituto Superior T\'ecnico,
Av. Rovisco Pais, 1049-001 Lisboa, Portugal 
}
\begin{abstract}
We apply the Bogoliubov-de Gennes equations to the confinement of a 
monopole-antimonopole pair in a superconductor.
This is related to the problem of a quark-antiquark pair bound by a
confining string, consisting of a colour-electric flux tube, dual to the 
magnetic vortex of type-II superconductors.
We study the confinement of the field lines
due to the superconducting state and calculate the effective potential
between the two monopoles. At short distances the potential is Coulombic
and at large distances the potential is linear, as previously determined
solving the Ginzburg-Landau equations. The magnetic field lines and the
string tension are also studied as a function of the temperature $T$.
Because we take into account the explicit fermionic degrees of freedom, 
this work may open new perspectives to the breaking of chiral symmetry 
or to colour superconductivity. 
\end{abstract}
\maketitle

\section{Introduction}

The confining string is an important candidate to solve the confinement problem
of quantum chromodynamics (QCD).
Right after QCD was proposed as the theory of strong interactions
\cite{Gross:1973id,Politzer:1973fx},
different ideas for confinement were developed. 
Nielsen and Olesen 
\cite{Nielsen:1973cs}
suggested that the magnetic flux tube vortex of type-II superconductors 
could be applied, after a dual transformation, to colour-electric flux tube strings
\cite{Goto:1971ce}
in QCD. Soon afterwards Nambu 
\cite{Nambu:1974zg},
't Hooft
\cite{'tHooft:1974qc},
and Mandelstam 
\cite{Mandelstam:1974vf}
developed the thesis that the open colour-electric string was able to confine 
quark-antiquark systems, where the 
quark-antiquark pair is dual to a Dirac
\cite{Dirac:1948um}
monopole-antimonopole pair.
This idea was further developed by
Baker, Ball and Zachariasen 
\cite{Baker:1984qh, Baker:1991bc}
and co-authors, who developed dual QCD, and also studied systems
similar to a monopole-antimonopole pair in a superconductor
in the Ginzburg-Landau framework. 
At the onset of QCD, lattice QCD was also developed by Wilson
\cite{Wilson:1974sk},
who showed analytically that the strong coupling limit
of QCD is equivalent to a string theory. The study of strings at the realistic
transition between the strong and weak coupling of Lattice QCD was further 
explored by several authors.
Bali
\cite{Bali:1992ab,Bali:1994de}
and Polikarkov
\cite{Ivanenko:1990xu,Polikarpov:1987yr}
and co-authors studied in detail the string formation between static
quark-antiquark sources in Lattice QCD.
Importantly, color superconductivity was also proposed to
exist at finite baryon density by Alford, Rajagopal and
Wilczek
\cite{Alford:1997zt}.

It is clear that QCD differs from type-II superconductors in
many details.
Baker, Ball and Zachariasen found that dual QCD is at the frontier
between type-I and type-II superconductivity.
The spin dependence of the quark-antiquark interaction,
unparallel to monopole-antimonopole interaction, was
also addressed by De Rujula, Georgi and Glashow, and
\cite{DeRujula:1975ge}
Henriques, Kellett and Moorhouse
\cite{Henriques:1976jd}
identified the role of the confining string in the spin 
dependent interactions.
Moreover Takahashi, Suganuma and co-authors
\cite{Takahashi:2000te,Takahashi:2002it}
also studied the three-legged string formation 
between three static quark sources, proposed to exist in Baryons.
It is clear that this needs a SU(3) symmetry, unparallel to type-I
or type-II superconductors. Okiharu, Suganuma and Takahashi
\cite{Okiharu:2004wy,Okiharu:2004ve},
further studied the tetraquark and pentaquark potentials.
The idea of bag model, separating the interior of the flux
tube from the remaining of the universe, was developed by
Chodos, Jaffe, Johnson, Thorn and Weisskopf
\cite{Chodos:1974je} ,
and this lead Jaffe and Johnson to propose the existence
of exotic hadrons, including glueballs, needing also a SU(3)
symmetry,
\cite{Jaffe:1975fd}.

Moreover, many problems in QCD confinement with strings remain 
to be solved. For instance in the limit of infinitely thin 
relativistic strings
\cite{Goddard:1973qh},
the world sheet swept out by a string in space time 
can only be quantized in 26 dimensions.
For example light quarks, which condense the vacuum with 
chiral symmetry spontaneously breaking $^3P0$ scalar 
quark-antiquark pairs described by Bicudo and Ribeiro
\cite{Bicudo:1989sh},
are not yet fully compatible with string confinement.
These problems, and other open problems of QCD confinement,
motivate very interesting and active investigations.

\begin{figure}
\includegraphics[width=0.75\textwidth]{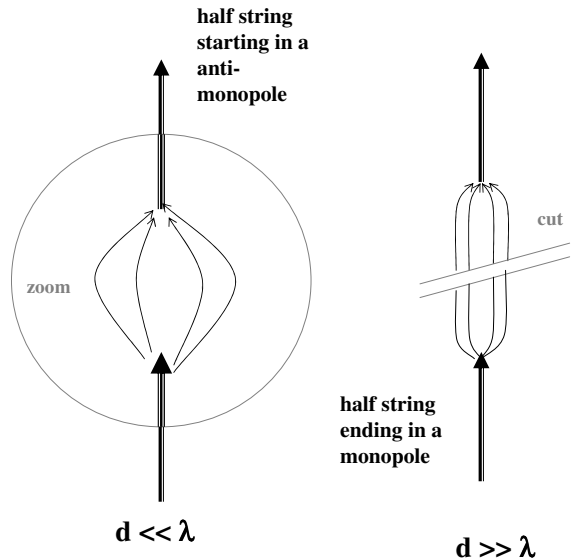}
\caption{\label{fig1} The magnetic field when the distance $d$
between the open ends of the two half strings is respectively much
smaller and much larger than $\lambda$.}
\end{figure}

Nevertheless the main inspiration of QCD confinement,
the monopole-antimonopole pair in a superconductor,
remains to be fully explored. Ball and Caticha
\cite{Ball:1987cf},
already studied the  monopole-antimonopole pair in 
Ginzburg-Landau type-II superconductor. 
The confinement of magnetic field lines in superconductors occurs
through an Anderson-Higgs mechanism such that the photons of the electromagnetic
field acquire a mass and are therefore exponentially damped in the superconductor.
This constitutes the Meissner effect discovered experimentally in 1933.
Indeed at the boundary between the exterior and the superconductor, the parallel
component of the external magnetic field is damped inside the superconductor
on a scale $\lambda$, the penetration length. 
When the distance $d$ between the monopole and the antimonopole 
is smaller than $\lambda$, the Coulomb interaction is dominant.
When $d>> \lambda$, the linear interaction, characterized by the
string tension $\sigma$, is preponderant.   
This is schematically shown in Fig. \ref{fig1}.
Another relevant scale in a superconductor
is the coherence length, defined as the distance over which the superconducting
order parameter (pair density) acquires its bulk value. The ratio of these two lengths
determines the magnetic properties of the superconductor: for $k=\lambda/\xi<1/\sqrt{2}$
the superconductor is of type-I, a complete Meissner effect is observed and the
magnetic field is completely expelled from the superconductor (perfect diamagnet)
over a distance defined by the penetration length.
If $k=\lambda/\xi>1/\sqrt{2}$ the superconductor is of type-II and the
magnetic field penetrates in the material through quantized vortices, whose density
increases until their cores (with size given by the coherence length) overlap
and the system becomes normal.

In this paper we apply the 
Bogoliubov-de Gennes (BdG) equations, enabling a microscopic
understanding of the role of electronic states in the confinement
of magnetic and electric fields in superconductors. This
is an inspiring model for QCD, where colour superconductivity 
and chiral symmetry breaking can both be addressed at the 
microscopic level of fermion (quark or antiquark) pairing.
We will be interested in the situation where two solenoids or magnetic
rods are inserted in a superconductor. In the limit of a very
thin solenoid, the open end of the solenoid is equivalent to a magnetic 
monopole plus a string. It is similar to a magnetic monopole and its 
associated Dirac string introduced to satisfy Maxwell's equation 
$\mathbf{\nabla} \cdot \mathbf{B}=0$.
This illustrates the presence of a heavy quark-antiquark pair 
in QCD. The BdG framework, applied to the monopole-antimonopole pair
is detailed in Section II. In section III we show the results
at $T=0$, and compare them with the Ball-Caticha paper.
In Section IV we show the results at $T\neq 0$. The finite
temperature and finite fermion density results of this
paper are interesting to illustrate what may happen in heavy ion 
collisions or in dense stars. In Section V we conclude.

\section{Bogoliubov de Gennes equations and the monopole field}

\subsection{Fermion pairing}

\begin{figure}
\includegraphics[width=0.4\textwidth]{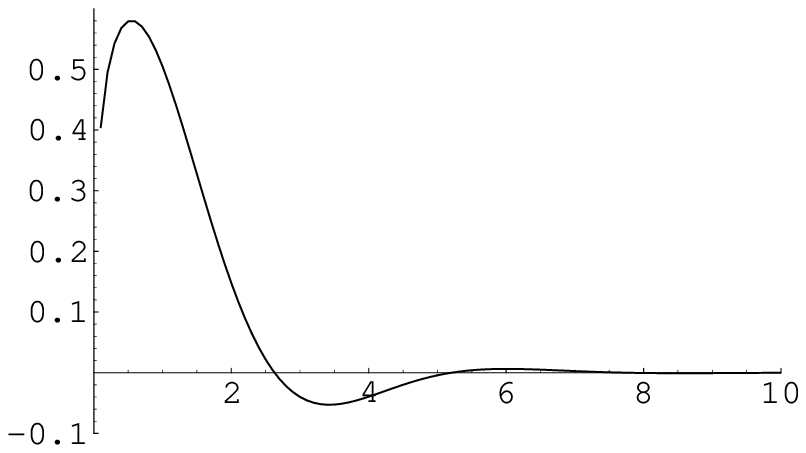}
\includegraphics[width=0.4\textwidth]{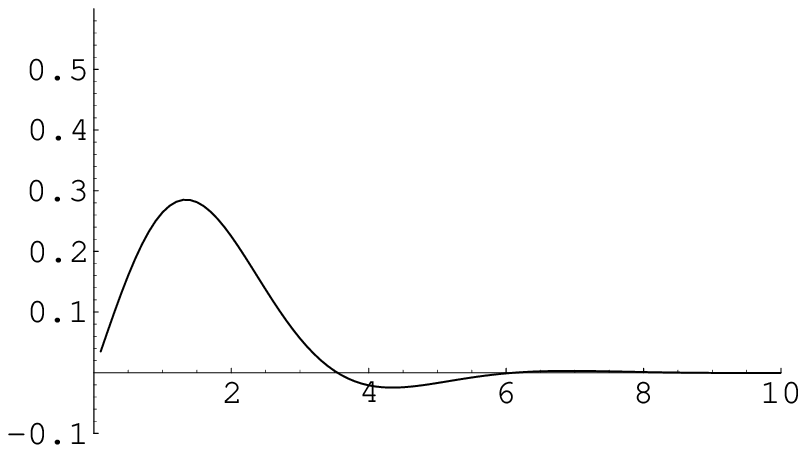}
\caption{\label{electronic} 
The $\rho$ dependence, in the equatorial plane ($z=0$), of the lowest 
electronic wavefunction. Notice that this is a localized 
wavefunction. We show in a) wavefunction $u$, and
in b) the wavefunction $v$
}
\end{figure}

Consider a fermionic system with an effective attractive interaction
between the particles defined by an Hamiltonian
\bea
H &=& \int d \mathbf{r} \sum_{\sigma} c^{\dagger}(\mathbf{r},\sigma)  \frac{ \left(\mathbf{p}-
\frac{e}{c} \mathbf{A} \right)^2}{2m} c(\mathbf{r},\sigma) \nonumber \\
&-& \frac{1}{2} g \int d \mathbf{r} \sum_{\sigma,\sigma'} c^{\dagger}(\mathbf{r},\sigma) 
c^{\dagger}(\mathbf{r},\sigma') c(\mathbf{r},\sigma') c(\mathbf{r},\sigma)
\eea
where $c^{\dagger}(\mathbf{r},\sigma)$ creates a fermion at site $\mathbf{r}$ with spin
$\sigma$. $\mathbf{A}$ is the vector potential.  
In BCS theory, this leads to fermion-fermion and antifermion-antifermion pairing.
Defining the field of a pair as $\Delta (\mathbf{r})$, and considering a conventional 
s-wave superconductor, the Hamiltonian is decoupled like
\bea
H_{\mbox{\tiny BCS}} &=& \int d \mathbf{r} \sum_{\sigma} c^{\dagger}(\mathbf{r},\sigma) 
\left[ \frac{ \left(\mathbf{p}- \frac{e}{c} \mathbf{A} \right)^2}{2m} -E_F \right] 
c(\mathbf{r},\sigma)
\\ \nonumber 
&+& \int d \mathbf{r} \left[ \Delta (\mathbf{r}) c^{\dagger}(\mathbf{r},\uparrow) 
c^{\dagger}(\mathbf{r},\downarrow) + \Delta^*(\mathbf{r}) c(\mathbf{r},\downarrow)
c(\mathbf{r},\uparrow) \right]
\eea
Here $E_F$ is the Fermi energy. The Hamiltonian is diagonalized through a canonical transformation
\bea
c(\mathbf{r},\uparrow) &=& \sum_i \left( u_i(\mathbf{r}) \gamma_{i,\uparrow} - v_i^*(\mathbf{r})
\gamma_{i,\downarrow}^{\dagger} \right) \nonumber \\
c(\mathbf{r},\downarrow) &=& \sum_i \left( u_i(\mathbf{r}) \gamma_{i,\downarrow} + 
v_i^*(\mathbf{r})
\gamma_{i,\uparrow}^{\dagger} \right) 
\eea
Here the fermionic operators $\gamma_{i,\sigma}$ annihilate a quasiparticle in the
level $i$ and with spin $\sigma$. 
Defining the diagonalized Hamiltonian as
\be
H = E_g + \sum_{i,\sigma} E_i \gamma_{i,\sigma}^{\dagger} \gamma_{i,\sigma}
\ee
where $E_g$ is the groundstate energy, the physical meaning of the new operators
is that they create the excitations of the system, and therefore we must restrict
ourselves to the positive energies. Using the commutation relations of the operators
with the Hamiltonian leads to the Bogoliubov-de Gennes equations (BdG)
\cite{book} for the amplitudes
$u(\mathbf{r})$ and $v(\mathbf{r})$:
\begin{equation}
\big( \frac{1}{2m}(\mathbf{p}-\frac{e}{c}\mathbf{A})^2 - E_F \big) u_i(\mathbf{r}) + 
\Delta(\mathbf{r}) v_i(\mathbf{r})  = E_i u_i(\mathbf{r})
\end{equation}
\begin{equation}
-\big( \frac{1}{2m}(\mathbf{p}+\frac{e}{c} \mathbf{A})^2 - E_F \big) v_i(\mathbf{r}) + 
\Delta^{*}(\mathbf{r}) u_i(\mathbf{r})  = E_i v_i(\mathbf{r})
\end{equation}
$\mathbf{A}(\mathbf{r})$ is the vector potential, $\Delta(\mathbf{r})$ is the pairing function
given by 
\begin{equation}
\Delta(\mathbf{r}) = g \sum_{0 \, < \, E_i \, \leq \,  \hbar \, {\omega}_{\scriptscriptstyle D} } u_i(\mathbf{r}) 
v_i^{*}(\mathbf{r})(1 - 2f(E_i)) \ .
\end{equation}
Here $f(E_i)$ is the Fermi-Dirac distribution given by
\be
f(E_i) = \frac{1}{\displaystyle e^{{}^{\frac{\scriptstyle  E_i}{{}_{\scriptstyle  k_B T}}}}+1}
\ee
where $T$ is the temperature, assumed
smaller than the critical temperature $T_c$, below which superconductivity arises. 
The pairing between the electrons
occurs on an energy scale $\hbar \omega_D$, called the Debye energy. In conventional
superconductors this is the cutoff energy provided by the phonon exchange
between electrons.
An example of electronic wavefunctions
$u$ and $v$ obtained in this paper is depicted in Fig. \ref{electronic}.
The vector potential is given by Maxwell's equations
\begin{equation}
\nabla \times \mathbf{B} = \nabla \times \nabla \times \mathbf{A}
 = \frac{4 \pi}{c}\mathbf{J}_{total}
\end{equation}
which, in the Coulomb gauge ( $\nabla . \mathbf{A} = 0$ ), is given by
\begin{equation}
\nabla^2 \mathbf{A} = - \frac{4\pi}{c} \mathbf{J}_{total}
\end{equation}
The current density originated in the supercurrents is obtained self-consistently by
\bea
\mathbf{J}(\mathbf{r}) &=& \frac{e\hbar}{im} \sum_i f(E_i) u_i^{*}(\mathbf{r})
( \nabla - \frac{ie}{\hbar c}\mathbf{A}(\mathbf{r}) ) u_i(\mathbf{r}) 
\\ \nonumber 
&+& (1-f(E_i)) v_i(\mathbf{r}) (\nabla -
\frac{ie}{\hbar c}\mathbf{A}(\mathbf{r}))v_i^{*}(\mathbf{r}) - c.c.
\eea

Let us take the order parameter in the form
\begin{equation}
\Delta(\mathbf{r}) = \Delta(\rho,z) e^{-in\varphi}
\end{equation}
This form describes a magnetic flux equal to $n$ flux quanta
($\Phi= n \phi_0 = n \frac{hc}{2e}$).
The wave functions $u_i$ and $v_i$ are expanded in a way similar to ref. \cite{Gygi}
\begin{equation}
u_i(\mathbf{r}) = \sum_{\mu,j,k} c_{\mu,j,k} \phi_{j,\mu-n/2,k} e^{i(\mu-n/2)\varphi}
\end{equation}
\begin{equation}
v_i(\mathbf{r}) = \sum_{\mu,j,k} d_{\mu,j,k} \phi_{j,\mu+n/2,k} e^{i(\mu+n/2)\varphi}
\end{equation}
where
\begin{equation}
\phi_{jmk} = \frac{\sqrt{2}}{RJ_{m+1}(\alpha_{jm})}J_{m}\left(\alpha_{jm}\frac{r}{R}\right)
\frac{e^{i\frac{2 \pi k z}{h}}}{\sqrt{h}}
\end{equation}
Here $\mu$ is an half-odd integer if $n$ is odd and an integer if $n$ is even. 
$J_{m}$ is a Bessel function.
The system is placed in a cylinder of radius $R$ and height $h$.
Given the azimuthal symmetry of the system
$\mathbf{A}$ does not depend on $\varphi$; therefore the Hamiltonian of the
BdG equations may be simultaneously diagonalized for each value of $\mu$.
It is therefore enough to diagonalize the matrix
\begin{equation}
\left( \begin{array}{cccc} 
		  T^{-} & \Delta \\
		  \Delta^T & T^{+} \\
		  \end{array}
		  \right)
\left( \begin{array}{c} 
		  c_i  \\
		  d_i \\
		  \end{array}
		  \right)		
= E_i
\left( \begin{array}{c} 
		  c_i  \\
		  d_i \\
		  \end{array}
		  \right)				   
\end{equation}
where
\bea
T_{j j' k k'}^{\pm} &=& \pm \frac{\hbar^2}{2m}( \alpha_{j\mu\pm n/2}^2 + 
\frac{k^2\pi^2}{h^2}) \delta_{jj'}\delta_{kk'} \nonumber \\ 
&-& (\mu \pm n/2) \frac{e}{\hbar c} I_1 
\pm \frac{e^2}{\hbar^2 c^2}I_2 - E_F
\eea
with
\bea
I_1 &=& \int_{0}^{h} dz \int_{0}^{R} \rho d\rho \nonumber \\
& & \phi_{j,\mu \pm n/2,k}(\rho,z) 
\frac{A_\varphi(\rho,z)}{\rho} \phi_{j',\mu \pm n/2,k'}(\rho,z)
\eea
\bea
I_2 &=& \int_{0}^{h} dz \int_{0}^{R} \rho d\rho \nonumber \\
& & \phi_{j,\mu \pm n/2,k}(\rho,z) A_\varphi(\rho,z)^2 
\phi_{j',\mu \pm n/2,k'}(\rho,z) 
\eea
We also obtain that
\bea
\Delta_{jj'kk'} &=& \int_{0}^{h} dz \int_{0}^{R} \rho d\rho \nonumber \\
& &  \phi_{j,\mu - n/2,k}(\rho,z) \Delta(\rho,z) 
\phi_{j',\mu + n/2,k'}(\rho,z) \nonumber \\
\eea

\subsection{Monopolar gauge field}

\begin{figure}
\begin{picture}(350,200)(0,0)
\put(-20,-320){
\includegraphics[width=0.75\textwidth]{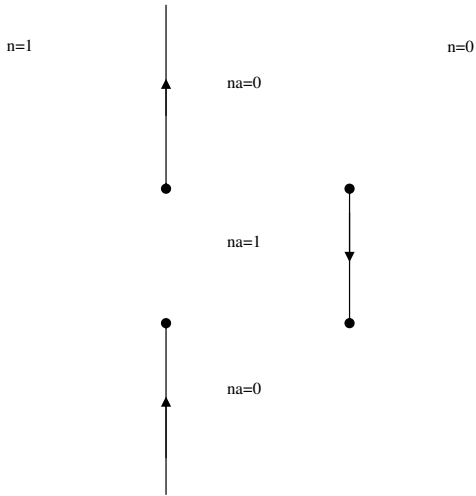}
}
\end{picture}
\caption{\label{fig3} 
Two choices of string orientation}
\end{figure}

The vector potential has two contributions one due to the magnetic monopoles, $\mathbf{A}^{ext}$
and the other due to the currents that are formed in the superconductor
, $\mathbf{a}$. Due to the linearity of the Maxwell equations we have that
\begin{equation}
\mathbf{A} = \mathbf{A}^{ext} + \mathbf{a}
\end{equation}
The contribution from the magnetic monopoles corresponding to a magnetic field
\be
\mathbf{B} = \frac{g_M}{4\pi} \frac{\mathbf{r}}{r^3}
\ee
where $g_M$ is the magnetic charge and $r$ the distance to the monopole,
is given by
\begin{equation}
\mathbf{A}^{ext} = -\frac{g_M}{4\pi} \frac{\rho}{r(r-z)} \mathbf{e}_\varphi
\end{equation}
This form has a singularity at the origin and on the line $(0,0,z)$ with $z > 0$.
We consider two magnetic monopoles one located at $\mathbf{r}_1 = (0,0,z_1)$,
and the other located at $\mathbf{r}_2 = (0,0,z_2)$. We choose $z_2 > z_1$,
and therefore we have singularities at $z > z_2$ and $z < z_1$. 
The potential is regularized taking $r \rightarrow R=\sqrt{r^2+\delta^2}$. This
is equivalent to replacing the magnetic monopole by a semi-infinite infinitely thin
solenoid corresponding to the magnetic charge and the Dirac string that carries
the magnetic flux from the infinity to the charge. The string guarantees that
$\nabla \cdot \mathbf{B}=0$.
We are interested in a situation where the two solenoids (or magnetic whiskers)
inserted in the superconductor originate from one monopole flux lines that 
penetrate the superconductor and are recovered in the other monopole.
In order for the
magnetic field lines to diverge from one monopole and converge in the other the
magnetic field created by the two monopoles has the form
\be
\mathbf{A}^{ext} = -\frac{g_{M}}{4\pi}\left\{  \frac{\rho}{r_1[r_1-(z-z_1)]} + 
\frac{\rho}{r_2[r_2+(z-z_2)] } \right\} \mathbf{e}_\varphi
\ee
with $r_i = \sqrt{\rho^2+(z-z_i)^2}$.

Close to each Dirac string and far from the monopoles the external vector potential is
approximately given by a term like
\begin{equation}
A_{\varphi}^{ext} = n_{ext} \frac{\hbar c}{2 |e| \rho}
\end{equation}
since we can relate $g_M$ with the total flux $\Phi$ coming out from a monopole
where $\Phi=n_{ext} \Phi_0$. Then it can be shown \cite{vorticity} that
the BdG equations can be recast in a form such that they only depend
on the difference $n_a=n-n_{ext}$ between the vorticity of the gap
function ($n$) and the vorticity of the external line ($n_{ext}$).
In the BdG equations we just have to replace $n\rightarrow n_a$ and
the effect of the external infinitely thin solenoid is taken into account
exactly if $n_{ext}$ is an integer. In our case the vector potential
is more complex. 

The choice of form for the vector potential implies that the Dirac strings
go from each monopole to infinity aligned with the $z$-axis in such a way that
the direction of flux goes from $-\infty$ to $+\infty$. Assuming that $g_M=\Phi_0$
we have an unit of flux along the $z$-axis, as shown in Fig. \ref{fig3}. The flux lines
diverge (converge) from (to) the south (north) monopoles. The flux due to the
field lines coming out (or in) of each monopole is equally divided $1/2$ ($-1/2$)
between the north and the south. In addition we have to take into account
the flux carried by the strings. Therefore considering any plane perpendicular
to the $z$-axis, the flux traversing it will contain a flux quantum. In the south
part the contributions add like: string ($+1$ flux), south monopole ($-1/2$) and
north monopole ($+1/2$). In the region between the monopoles where a flux tube confined
by the superconductor will appear we have: south monopole ($+1/2$) and
north monopole ($+1/2$). Notice that with this choice there is no string traversing
an horizontal plane when $z_1<z<z_2$. In the north part the contributions are such:
string ($+1$ flux), south monopole ($+1/2$) and north monopole ($-1/2$). Therefore
it always adds to a unit of flux.

We can as well consider that the Dirac string
has a finite extension and goes from the north monopole to the south monopole, as
shown as well in Fig. \ref{fig3}. In this case it is easy to see that the total flux is zero.
It seems therefore natural to look for solutions where in the first case $n=1$ and
in the second case $n=0$, reflecting the different total magnetic fluxes.

In the first case it is easy to see that the effect of the strings is such that,
in an approximate way, in the region $z<z_1$ close to the string and far from the monopole,
the BdG equations are approximately described by the equations where the effect of the
external vector potential is equivalent to $n_a=0$, since we choose $n=1$ and the singular
contribution from the string is described by $n_{ext}=1$. A similar analysis
for $z>z_2$ leads to the same conclusion. Between the monopoles a finite width flux tube
forms. As shown in ref. \cite{vorticity} these field lines are not singular and
the vorticity of the equations is not affected: there is no string in this region.
Therefore $n_a=n-n_{ext}=1-0=1$. A similar analysis for the second string configuration
leads to a situation where for $z<z_1$ and $z>z_2$ we have an effective $n_a=0$ and
between the monopoles we have $n_a=1$. In this case there is a string between the
monopoles (with $n_{ext}=-1$). Therefore the two configurations are equivalent as
expected \cite{Dirac:1948um}.

\subsection{Numerical method}

The considerations above only apply to the singular part of the vector potential
due to the strings. The singular part effectively changes the set of basis functions
with the replacement $\mu \pm n/2$ to $\mu \pm n_a/2$. Therefore to take advantage
of this basis we just add and subtract to the external vector potential
the limiting singular expressions due to the strings. The remaining part of
the vector potential is regular. It has been shown by several authors that in this
case the contribution of the terms involving the vector potential in the BdG
equations is negligible or at least very small.
In the following we neglect $I_1$ and $I_2$.
We may therefore divide the system into two regions, using the inversion
symmetry around $z=0$. For $|z|<|z_i|$ we solve the BdG equations taking a basis
with $n_a=1$ (equivalent to the problem of a singly quantized vortex line with
no external vector potential) and for $|z|>|z_i|$ we use a basis with $n_a=0$.
We should note that, as shown in ref. [28], we have to consider large
systems and consequently a large basis of functions has to be used to properly obtain
the exponential decay of the various physical quantities. Otherwise,
the exponential regime is not reached. This implies that solving the full
$3d$ problem requires very large matrices to be diagonalized. Therefore
we have to resort to this approximation for the $z$ dependence.

In this approximation the eigenfunctions
of the BdG Hamiltonian in each region are plane waves along $z$. 
Therefore the potential $\Delta$ does not depend on $z$ and the BdG Hamiltonian
is independent of $z$. It may be block diagonalized in terms of the momentum
along the $z$ direction (indexed by $k$). In this way the components of the Hamiltonian are
reduced to
\begin{equation}
T_{j j'}^{\pm} = \pm \frac{\hbar^2}{2m}
\left( \frac{\alpha_{j\mu\pm n_{\scriptstyle a}/2}^2}{R^2} + \frac{4k^2\pi^2}{h^2}\right) 
\delta_{jj'} \mp E_F
\end{equation}
\begin{equation}
\Delta_{jj'} = \int_{0}^{R} \phi_{j,\mu - n_{\scriptstyle a}/2}(\rho) \Delta(\rho) 
\phi_{j',\mu + n_{\scriptstyle a}/2}(\rho) \rho d\rho
\end{equation}
It is important to note that the symmetry of the BdG equations
$u_i(\mathbf{r}) \rightarrow v_i^{*}(\mathbf{r})$,
$v_i(\mathbf{r}) \rightarrow -u_i^{*}(\mathbf{r})$
and $E_i \rightarrow - E_i$
allows to reduce the solution to the positive values of $\mu$.
We obtain the eigenvectors and eigenvalues for negative values of $\mu$ using the
above symmetry.

\begin{figure}
\includegraphics[width=0.15\textwidth]{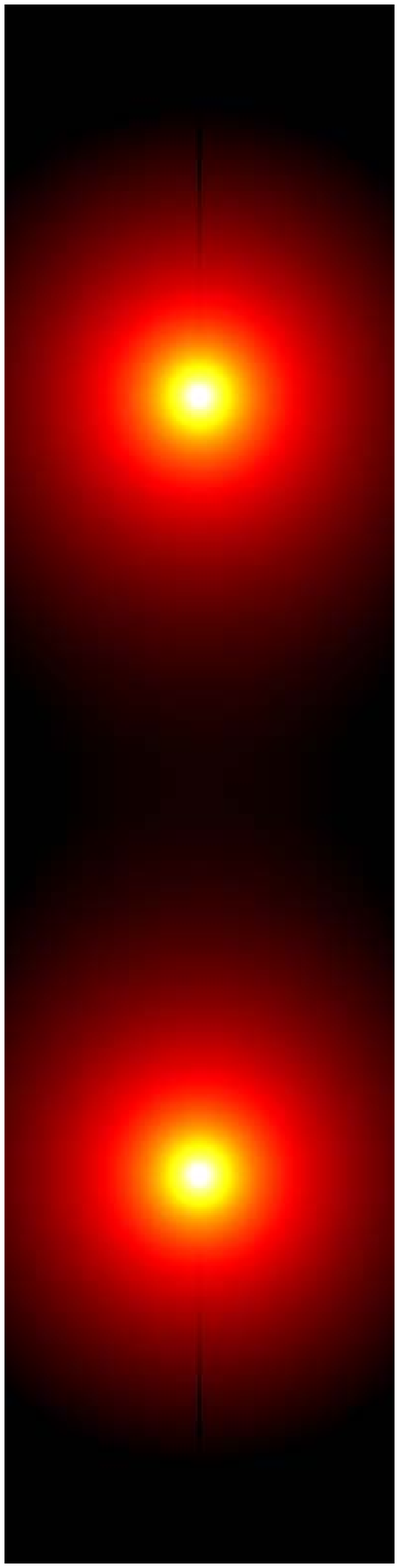}
\includegraphics[width=0.15\textwidth]{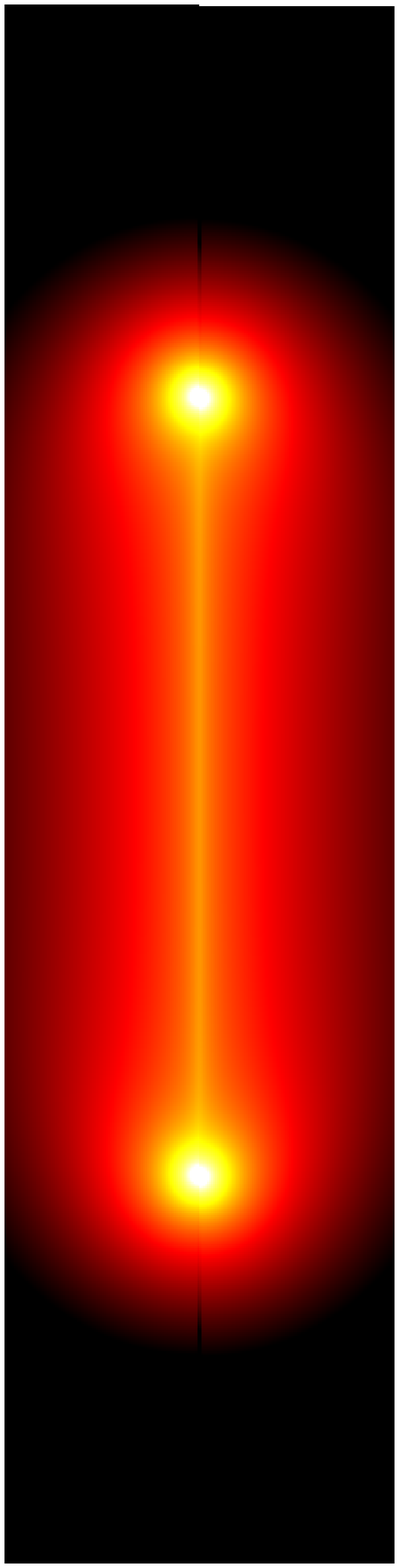}
\includegraphics[width=0.15\textwidth]{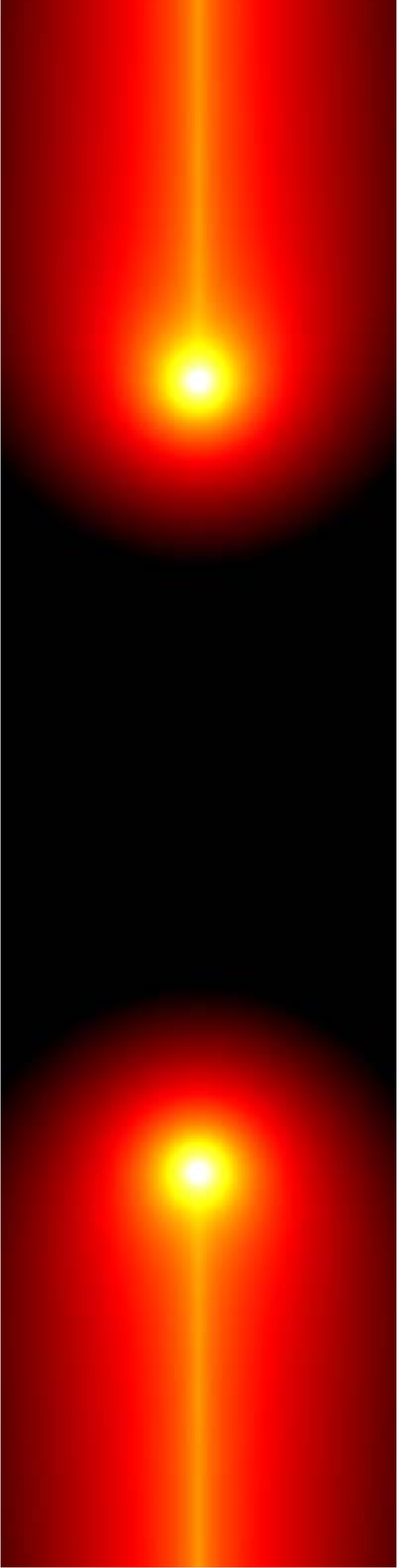}
\includegraphics[width=0.465\textwidth]{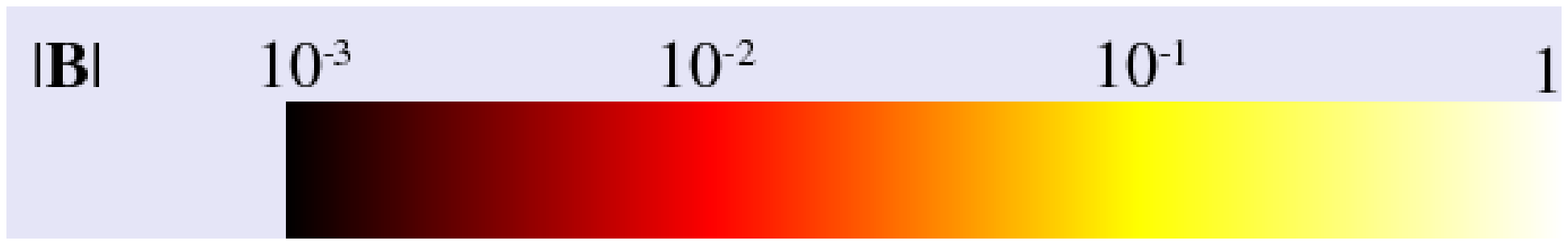}
\caption{\label{fig4} (colour online)
Density plot for $|\mathbf{B}|$ for $d=480$.
The panels correspond from left to right to no superconductor in an external
field, the total field for $n=1$ and for $n=0$, respectively.}
\end{figure}

\begin{figure*}
\includegraphics[width=0.4\textwidth]{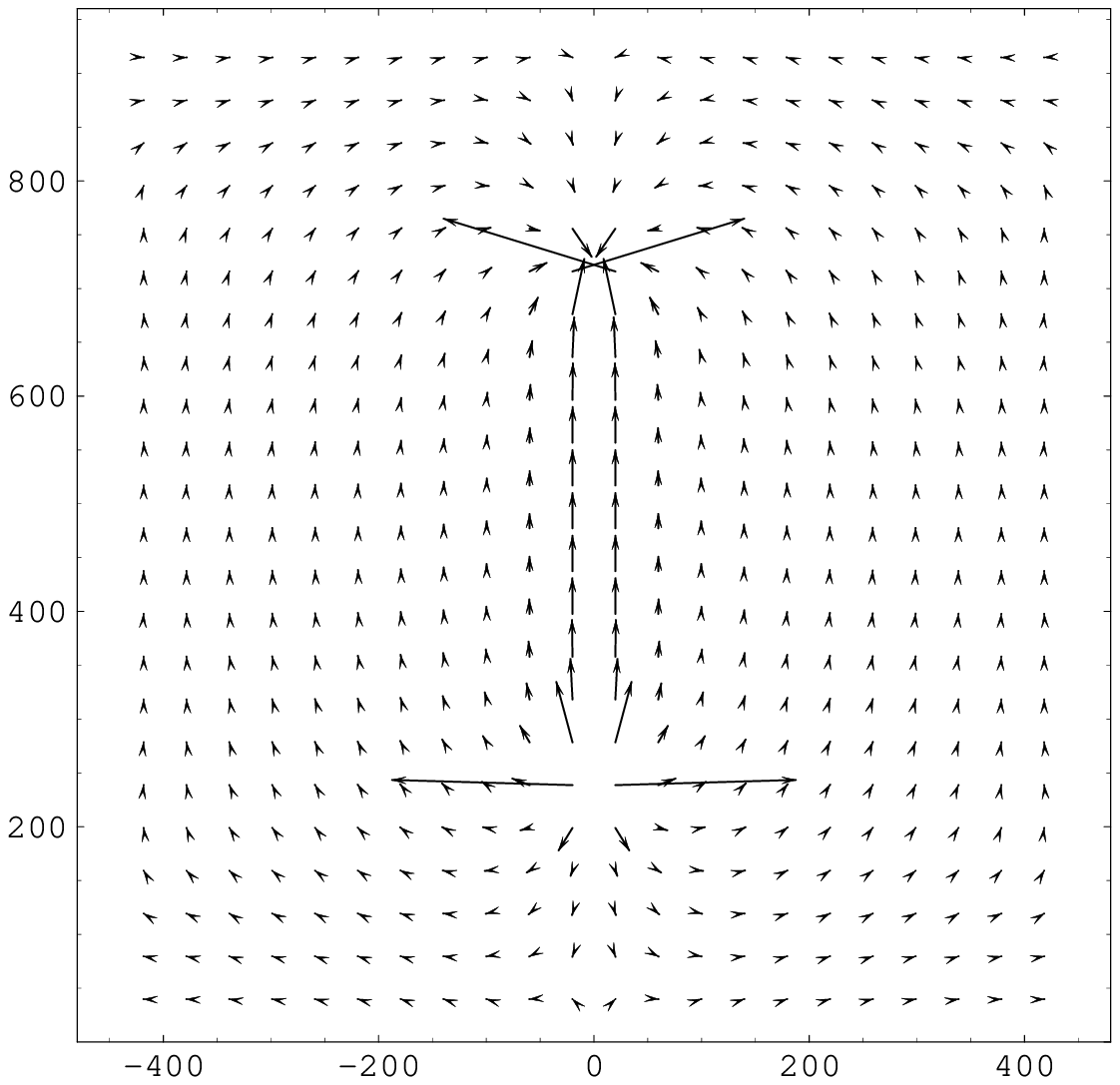}
\hspace{0.25cm}
\includegraphics[width=0.4\textwidth]{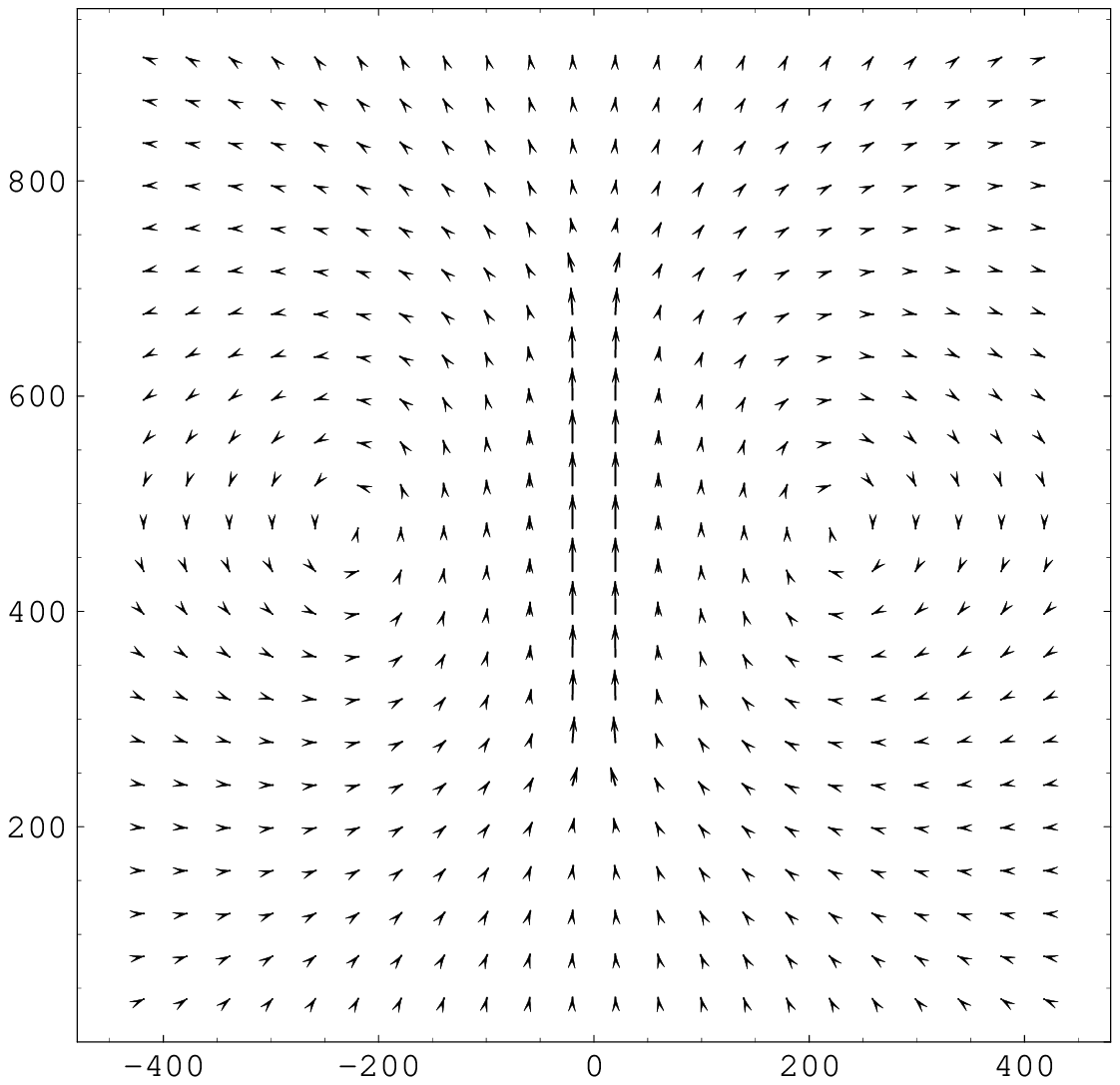}
\includegraphics[width=0.4\textwidth]{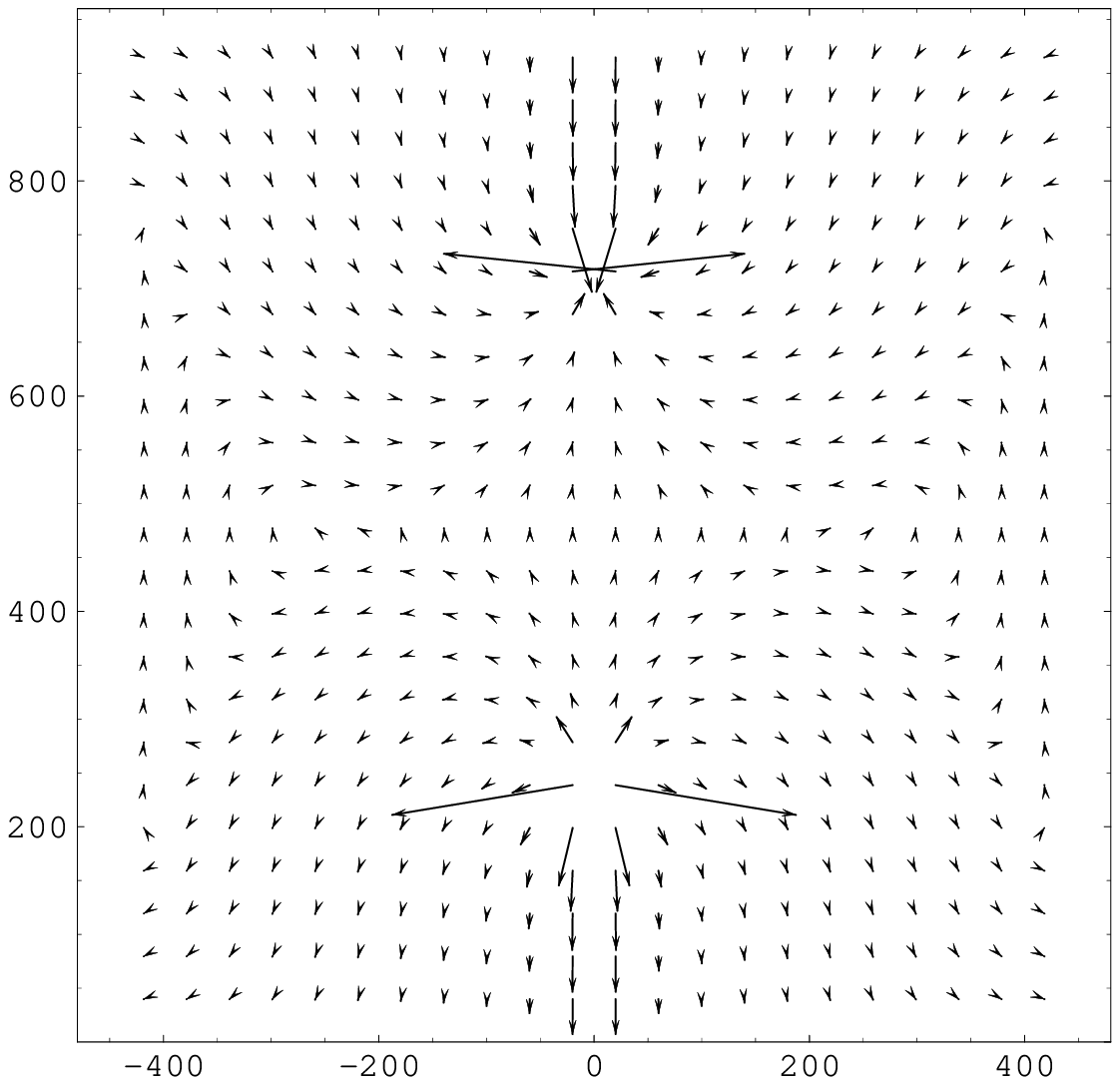}
\hspace{0.25cm}
\includegraphics[width=0.4\textwidth]{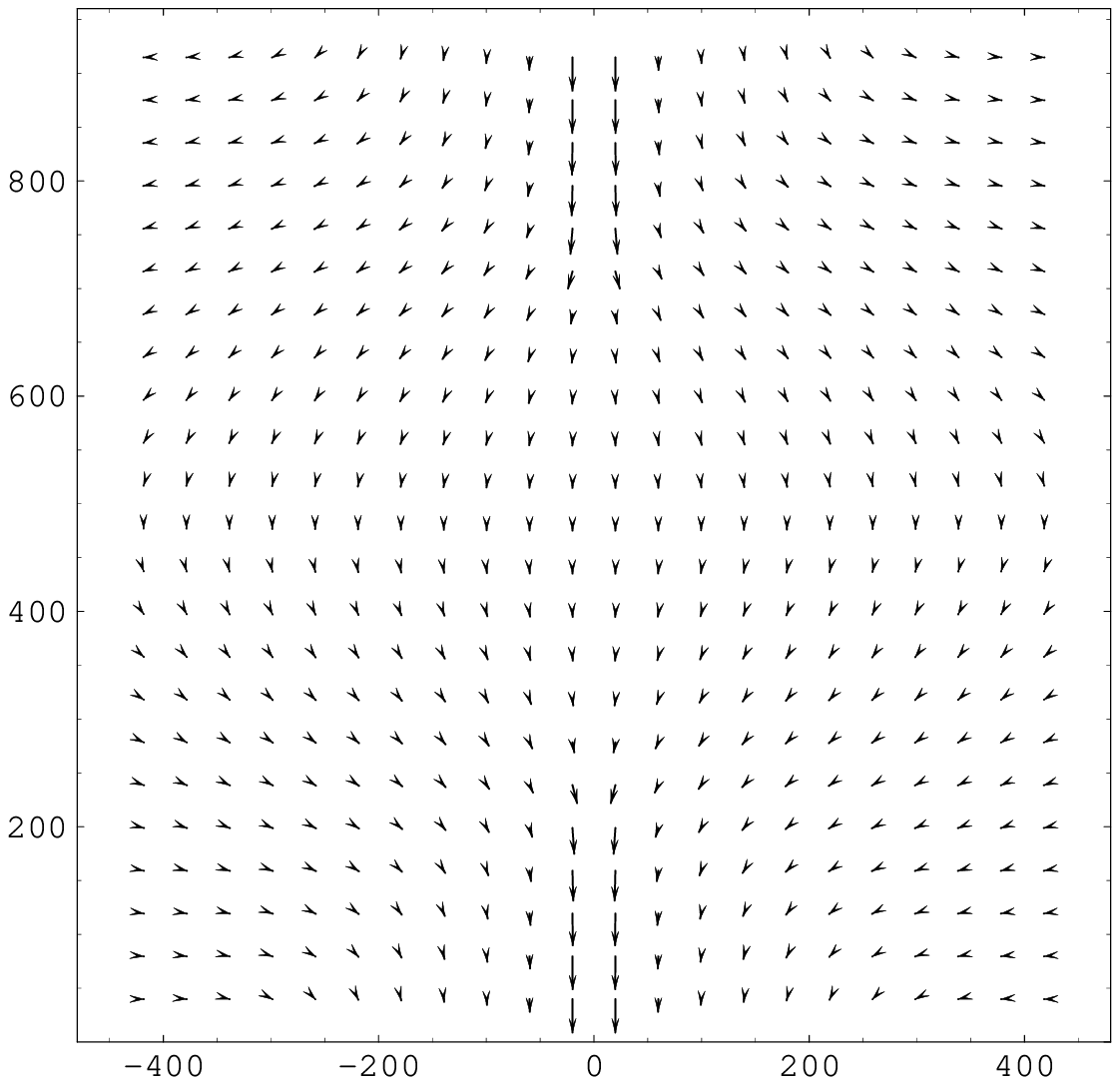}
\caption{\label{fig5} 
Magnetic field profile for the total field (left panels) and the internal field (right
panels). We consider the cases
of $n=1$ (upper panels) and $n=0$ (lower panels). 
Please note that the magnitude of the field is not visible here when
it is smaller than the size of the arrows. This information is
contained in Fig. \ref{fig4}. The size of the arrows is chosen for better visualization
of the field direction at each point.
}
\end{figure*}

We may find the internal vector potential solving Poisson's equation
$\nabla^2\mathbf{a} = -\frac{4\pi}{c}\mathbf{J}_s$.
Defining $a_{\varphi} = F(\rho) / \rho$, Posisson's equation
reduces in cylindrical coordinates to
\begin{equation}
\frac{\partial^2 F}{\partial \rho^2} - \frac{1}{\rho} 
\frac{\partial F}{\partial \rho} + \frac{\partial^2 F}{\partial z^2} = 
-\frac{4\pi}{c}J_\varphi \rho
\end{equation}
Decomposing the current as
\begin{equation}
-\frac{4\pi}{c}J_\varphi \rho = K(\rho,z) + \beta(\rho,z) F(\rho,z)
\end{equation}
with
\bea
& &K(\rho, z) = -\frac{4\pi}{c} \sum_{i} f(E_i) |u_i|^2 (\mu-n_a/2) 
\\ \nonumber
&-& (1-f(E_i)) |v_i|^2 (\mu+n_a/2) + \beta(\rho) \left( A^{ext}-a \right) \rho \ ,
\eea
\begin{equation}
\beta(\rho) = -\frac{4\pi}{c} \frac{|e|}{\hbar c}\sum_{i} f(E_i) |u_i|^2 + ( 1 - f(E_i))  
|v_i|^2   \ ,
\end{equation}
we get
\begin{equation}
\frac{\partial^2 F}{\partial \rho^2} - \frac{1}{\rho} 
\frac{\partial F}{\partial \rho} + \frac{\partial^2 F}{\partial z^2  }
=
K(\rho, z) + \beta(\rho) F(\rho) \ .
\end{equation}
These equations may be solved as described in ref. \cite{vorticity}.

\section{Results for $T=0$}

The parameters used are $\epsilon_F=0.5, \omega_D=0.25, g/\pi=1$ in atomic units.
We should note
that the relevant parameter is the ratio to the critical temperature.
For these parameters the critical temperature is of the order of $0.06$.
We consider the two monopoles in a system with height $960$ and radius $480$.
The distance between the monopoles, $d$, is varied. 
Let us consider that the strings carry a quantum of flux.

In Fig. \ref{fig4}
we compare the absolute value of the total magnetic field for the cases of no 
superconductor, and in the
superconducting phase considering a $n=1$ solution and comparing to the case of $n=0$.
The case of no-superconductor
is the usual magnetic dipole. 
Notice that the $n=1$ and the $n=0$ cases also correspond to two different configurations 
of confinement. The magnetic vortex provides a linear confining potential. In the $n=1$ 
configuration the monopole is attracted by the antimonopole, this is the standard picture of
confinement, say in a mesonic quark-antiquark system in QCD. In the $n=0$ configuration
the monopole is attracted by the boundary of the superconductor, not by the antimonopole.
The $n=0$ case simulates a situation
where the magnetic solenoids are being inserted from the border of the material. When 
the solenoids are far enough, it is favourable for the flux lines to emerge from the material.
As the monopoles get close together it is more favourable to establish a $n=1$ flux
tube between the monopoles.

Taking the solution of $n=1$ we look for a solution
where the flux is quantized and equal to a quantum of flux. 
Far from either monopole and close to the strings the magnetic field is small.
In the case of $n=0$ there is no flux through the system and the flux
lines leave the superconductor from the monopoles to the border of the system.
In Fig. \ref{fig5} we show the total and internal magnetic fields for both
$n=1$ and $n=0$. This figure provides complementary information with respect
to Fig. \ref{fig4}. In the case of $n=1$ the presence of the flux tube between the
monopoles is evident.  
In the case of $n=0$ the field lines avoid the central region except along the borders,
as expected. The diamagnetic behaviour of the internal field, obtained subtracting
the external field from the total field is clearly displayed in the right panels
of Fig. \ref{fig5}. In the case of $n=1$ the internal field close to the monopoles
is not strong enough to compensate the external field. Note however that
along the flux tube the internal field has the same orientation as the
total field. Interestingly the internal field in the vicinity of the
$z=0$ region and far from the axis winds clockwise and counter-clockwise.
In the case of $n=0$ once again along the flux tubes the internal field
has the same orientation as the total field. Between the monopoles the
internal field opposes the total field, as expected.

\begin{figure}
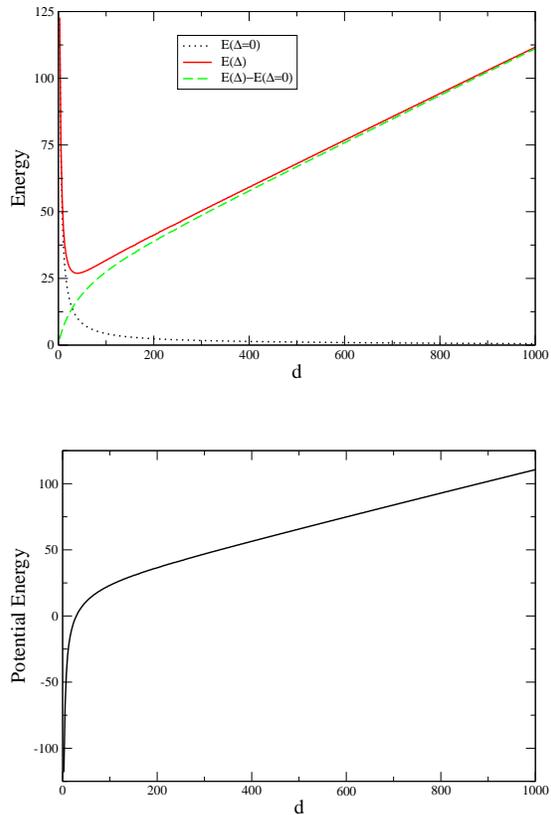

\begin{picture}(350,310)(0,0)
\put(0,160){
\includegraphics[width=0.4\textwidth]{fig6a.eps}
}
\put(0,-5){
\includegraphics[width=0.4\textwidth]{fig6b.eps}
}
\end{picture}
\caption{\label{fig6} (colour online)
a) Comparison of the total magnetic energy in the normal phase and in the
superconducting phase. Also, we present the energy difference
between the superconducting phase and the normal phase.
b) Energy difference between the superconductor and the normal phase
as a function of the distance between the monopoles plus the Coulomb
term.}
\end{figure}

Clearly it is interesting, in the analogy of the Anderson-Higgs mechanism to the confinement
of quarks, to determine the confinement potential between the two monopoles. Taking the
analogy of the monopoles with the quarks in QCD the fact that these are linearly confined
at large distances should translate in the superconductor to a linear potential energy
when the monopoles are moved apart. Indeed in Fig. \ref{fig6} we show the magnetic energy
of the system,
\be
U = \int \frac{|\mathbf{B}|^2}{8\pi} dV \ \ ,
\ee
as a function of the distance between the monopoles. The behaviour is
similar to the one proposed for the confinement problem in QCD. We show the difference
in energy between the normal system and the superconductor as a function of $d$.

Interestingly, the solenoid-antisolenoid magnetic
energy includes a subtle point. To illustrate it
let us now perform an analytical computation
of the energy of the solenoid-antisolenoid pair 
in the normal (with no superconductor) phase. 
The magnetic field $\mathbf{B}$ can be decomposed in the sum
of the field created by the north antisolenoid 1
and the south solenoid 2.
In the limit of infinitely thin solenoids,
each with a quantum of magnetic flux,
the respective fields are equivalent to a monopole
field plus a string field, 
\bea
{\bf B}_1 &=& { -g_m \over 4 \pi |r-r_0|^2} \widehat{r-r_0} - g_m 
\delta(x)\delta(y)\theta(z-z_0)(-\hat e_z) \ ,
\nonumber \\
{\bf B}_2 &=& { + g_m \over 4 \pi |r+r_0|^2} \widehat{r+r_0} + g_m 
\delta(x)\delta(y)\theta(-z-z_0)(\hat e_z) \ .
\nonumber \\
\eea
The $B_1^2+B_2^2$ contribution is infinite but it is also constant and independent
of the distance. The relevant non-constant term comes from the mixed
term. Then we get two different contributions for the energy.
The energy of the mixed monopole-antimonopole term is similar
to the usual electrostatic Coulomb energy,
\be
E_{mono-antimono}=- { g_m^2 \over 4 \pi |2 z_0|}  \ .
\label{eq:monoantimono}
\ee
However the energy also includes the monopole-string terms 
${{\bf B}_1}_{monopole} \cdot {{\bf B}_2}_{string}+
{{\bf B}_1}_{string} \cdot {{\bf B}_2}_{monopole}$. An inspection of the directions 
of the magnetic fields shows that these two terms don't cancel, in fact they
contribute equally. We get,
\bea
2 \, E_{string-antimono} &=& 2  \int_{z_0}^\infty dz \, 
{ + g_m \over 4 \pi (z+ z_0)^2} ( \hat e_z \cdot \hat e_z ) g_m
\nonumber \\
\nonumber \\
&=& 2 {g_m^2 \over 4 \pi} \left[  - 1 \over z+ z_0 \right]_{z_0}^\infty
\nonumber \\
&=& 2 {g_m^2 \over 4 \pi |2 z_0|} \ .
\label{eq:stringmono}
\eea  
Summing the two contributions the total energy is,
\be
U=+ { g_m^2 \over 4 \pi |2 z_0|}  + \mbox{ const.} \ \ \ .
\label{eq:totalwrongsign}
\ee
Notice that the final sign in eq. (\ref{eq:totalwrongsign}) is opposed 
to the normal coulomb energy of eq. (\ref{eq:monoantimono}).
This occurs because any movement of the solenoid in the magnetic
field of the other solenoid affects its magnetization and therefore 
changes its internal energy. In the present case, the magnetization 
of both solenoids has to be maintained constant, and this requires
an extra energy source, say a battery, thus affecting the energy 
balance of the system. Nevertheless the solenoid-antisolenoid 
magnetic force is attractive, and quite similar to the charge-anticharge 
electrostatic force.

\begin{figure}
\includegraphics[width=0.115\textwidth]{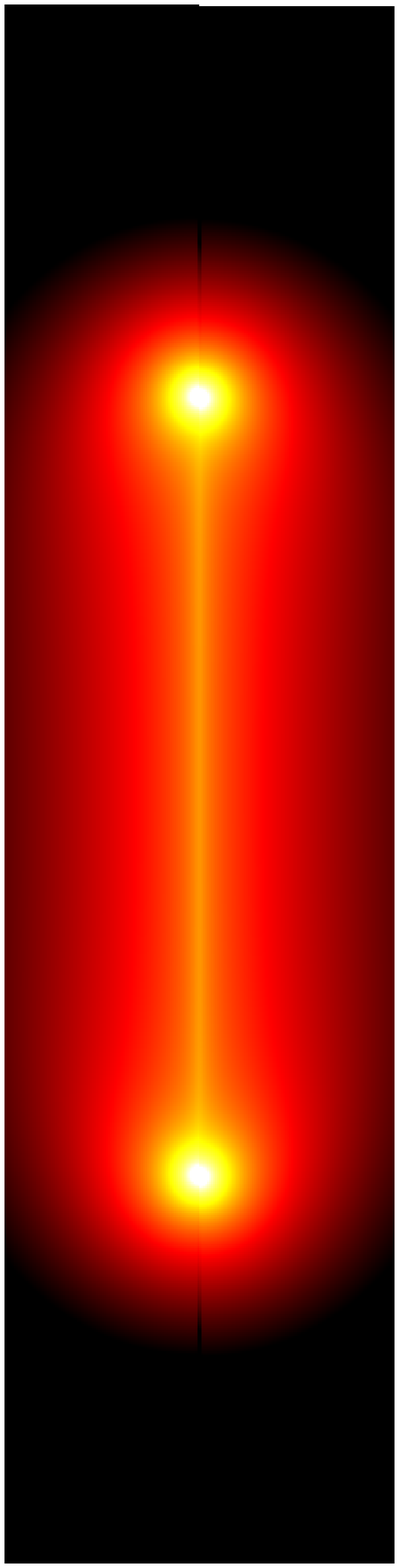}
\includegraphics[width=0.115\textwidth]{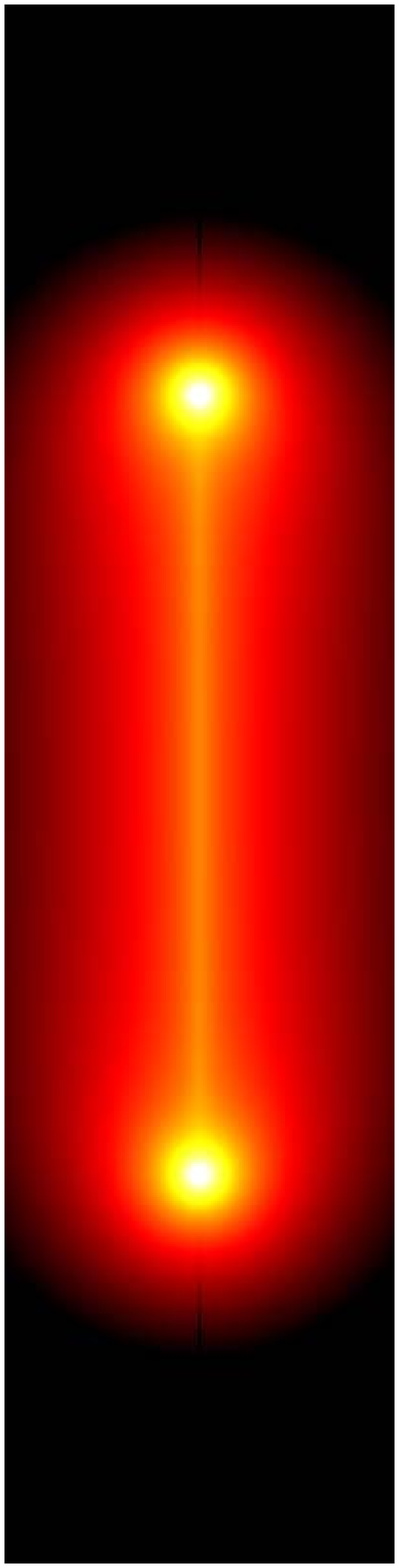}
\includegraphics[width=0.115\textwidth]{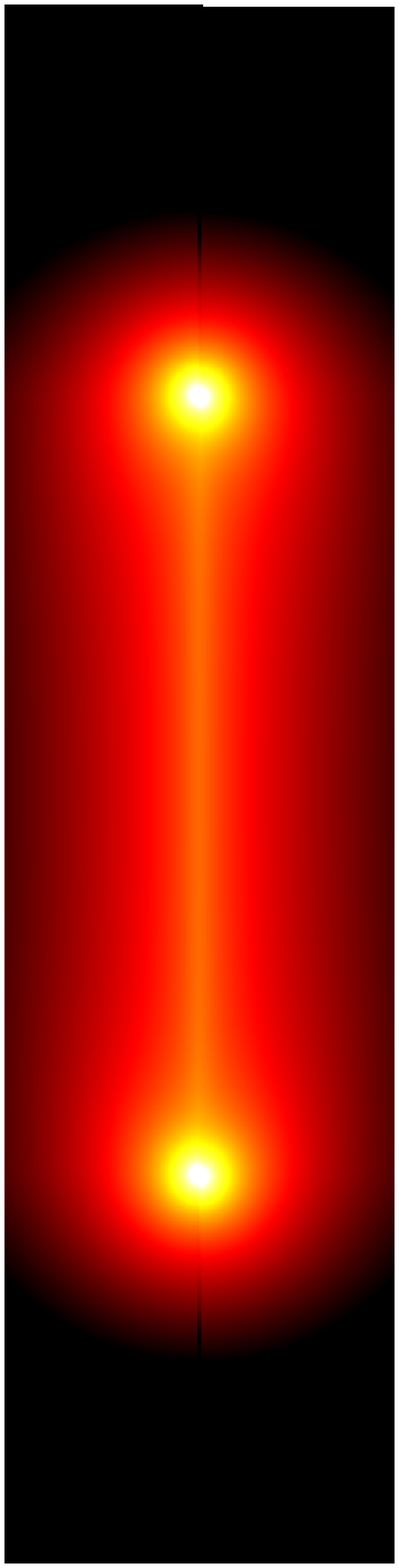} 
\includegraphics[width=0.115\textwidth]{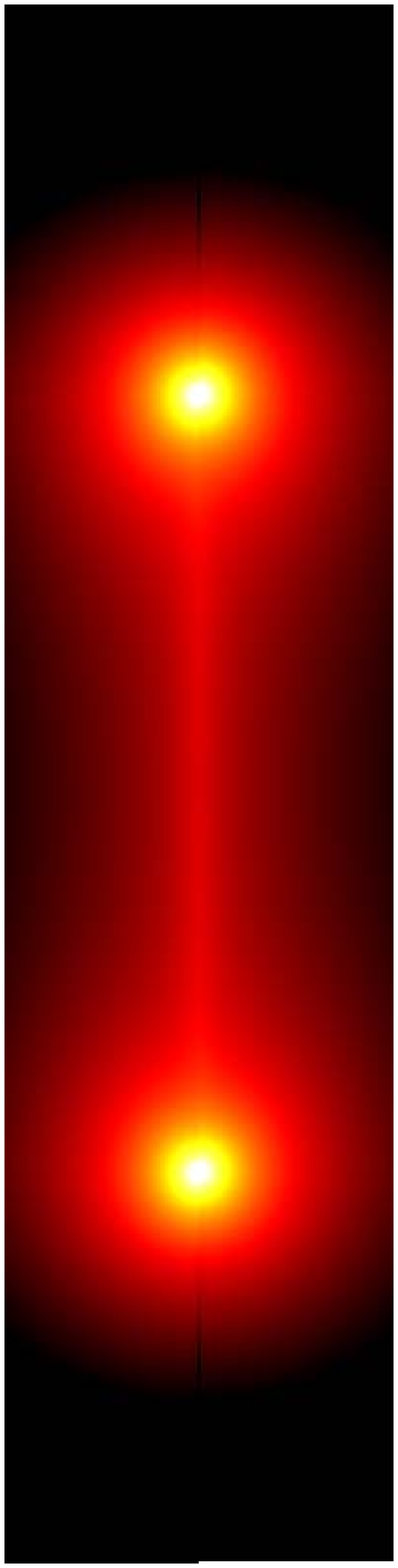} 
\caption{\label{fig7} (colour online)
Effect of temperature on the magnitude of the total 
magnetic field for $n=1$. The temperatures are, from left to right, $T=10^{-4},0.02,0.03
,0.04$, respectively.
For the density scale see Fig. \ref{fig4}.
}
\end{figure}

\begin{figure}
\includegraphics[width=0.115\textwidth]{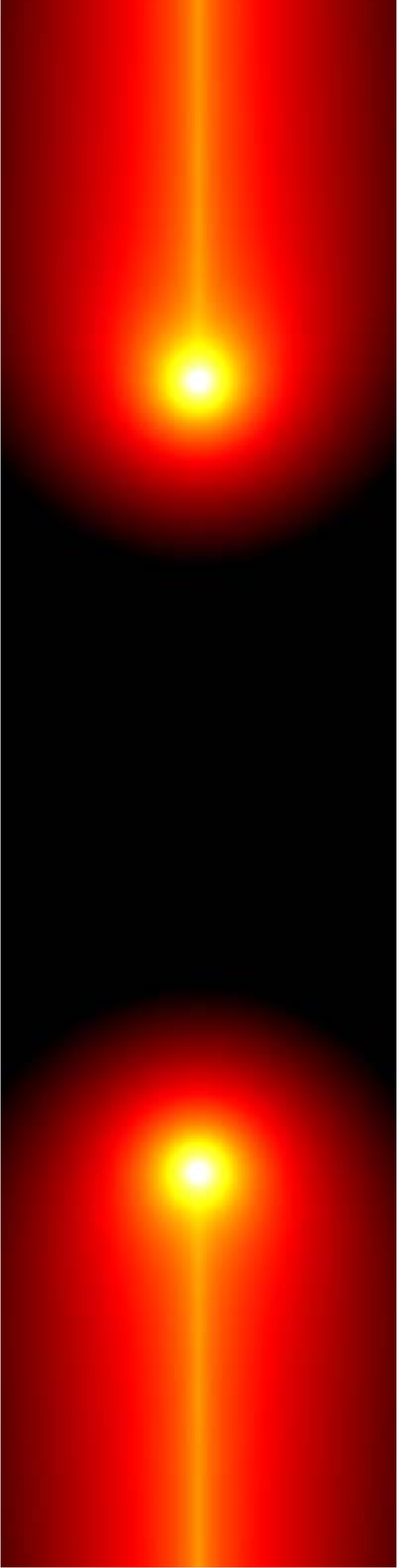}
\includegraphics[width=0.115\textwidth]{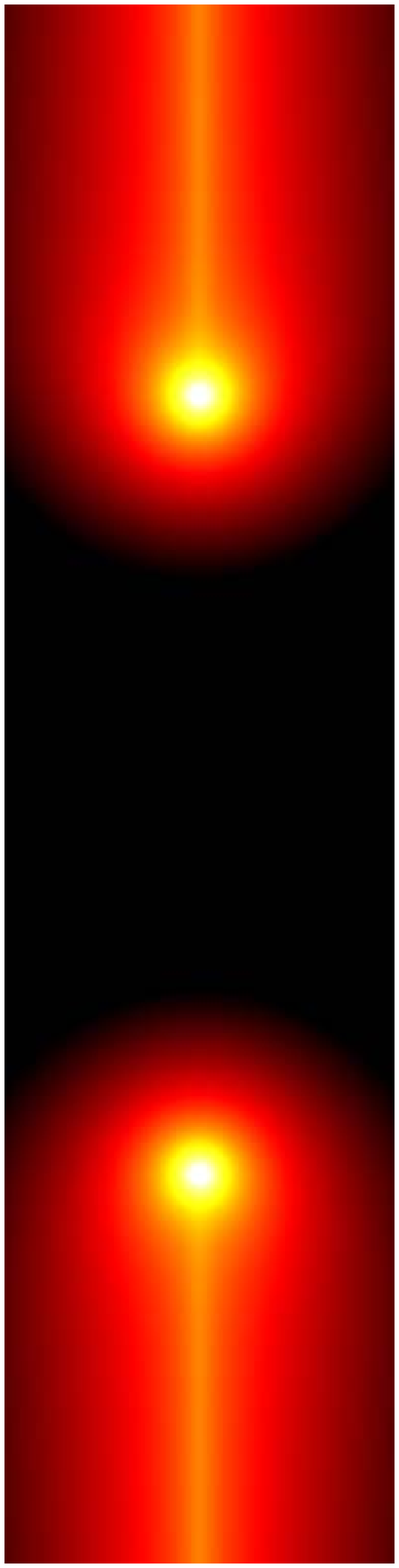}
\includegraphics[width=0.115\textwidth]{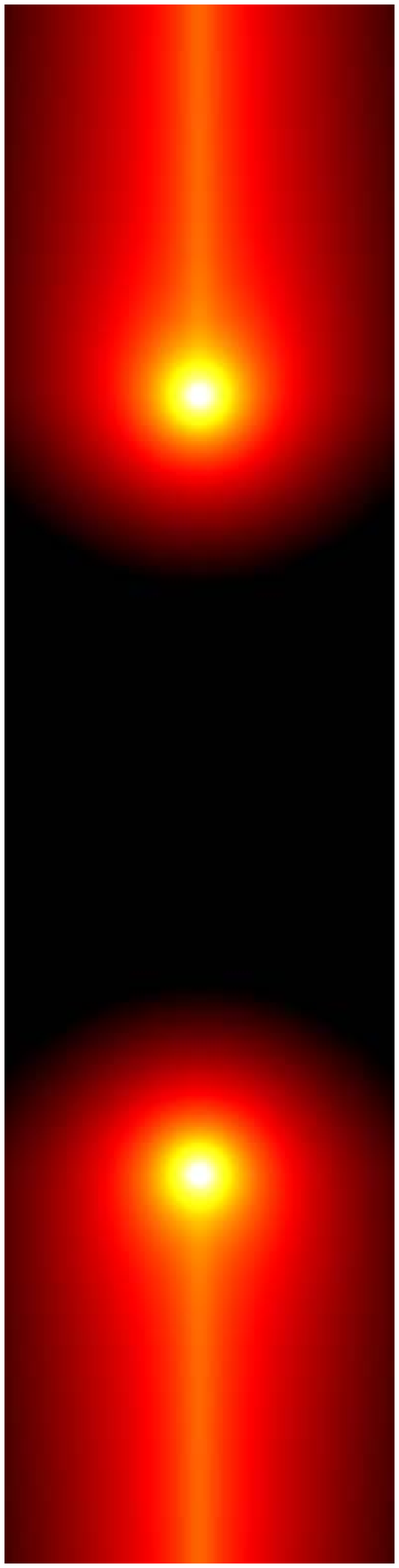} 
\includegraphics[width=0.115\textwidth]{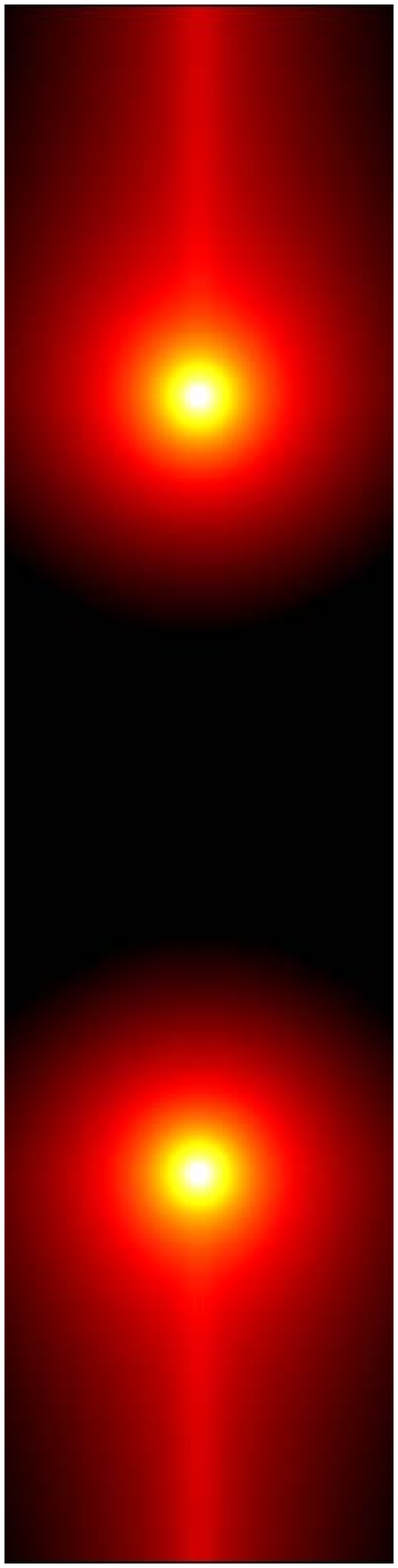} 
\caption{\label{fig8} (colour online)
Effect of temperature on the magnitude of the total 
magnetic field for $n=0$. The temperatures are, from left to right, $T=10^{-4},0.02,0.03
,0.04$, respectively.
For the density scale see Fig. \ref{fig4}.
}
\end{figure}

Indeed in Fig. \ref{fig6} a) we depict the magnetic energy
of the solenoid-antisolenoid system both in the normal
phase (no superconductor) and in the superconductor phase.
Although our solenoids are thinner than both the coherence 
length and the penetration length, they have a finite width,
and we are able to account for the inside (string-like) 
magnetic field created by the solenoids. 
Notice that the $1\over r$ component of the magnetic
energy is decreasing. To isolate
the linear potential we also show in Fig. \ref{fig6} a) the
difference between the magnetic energy in the two phases.
To get the monopole-antimonopole interaction, independently
of the string internal energy, we also add to the 
linear interaction of Fig. \ref{fig6} a) the monopole-antimonopole
Coulomb interaction of eq. (\ref{eq:monoantimono}).
This produces the potential depicted in \ref{fig6} b), 
similar to the linear+Coulomb potential of QCD.

\section{Results for $T \neq 0$}

\begin{figure*}
\subfigure{\includegraphics[width=0.3\textwidth
]{fig9d.eps}}
\hspace{0.5cm}
\subfigure{\includegraphics[width=0.3\textwidth
]{fig9e.eps}}
\hspace{0.5cm}
\subfigure{\includegraphics[width=0.3\textwidth
]{fig9f.eps}} 
\end{figure*}
\vspace{0.5cm}
\begin{figure*}
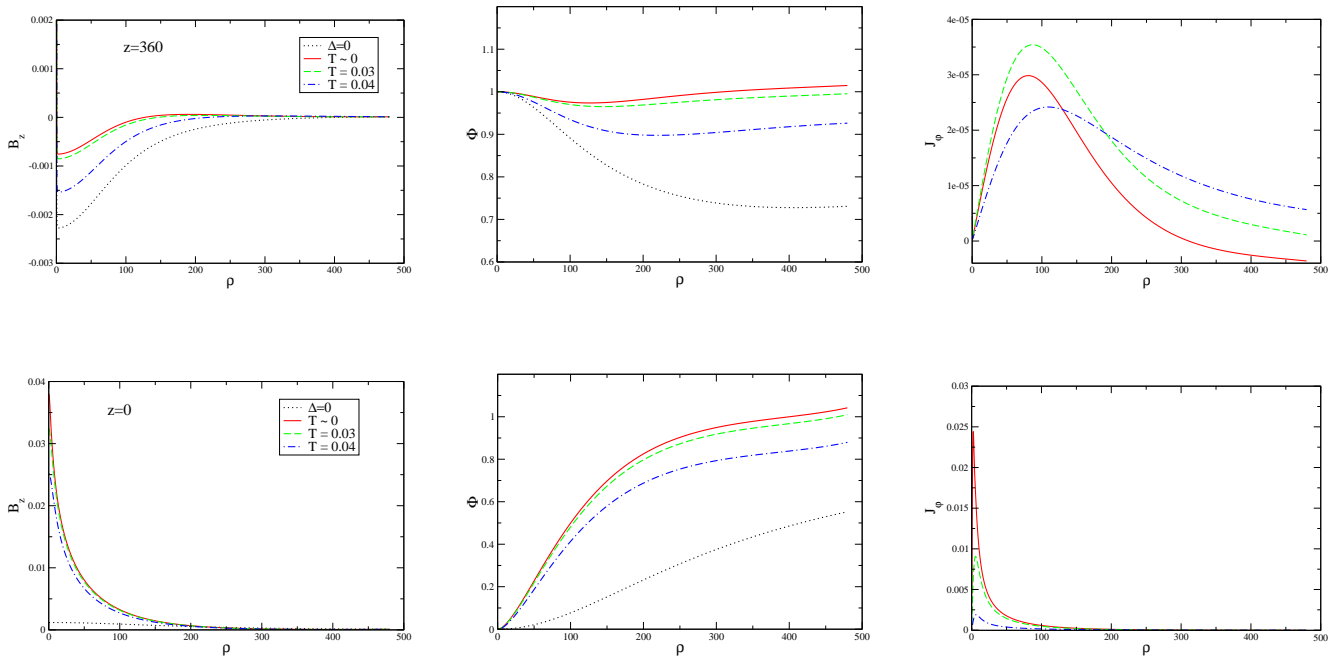

\subfigure{\includegraphics[width=0.3\textwidth
]{fig9a.eps}}
\hspace{0.5cm}
\subfigure{\includegraphics[width=0.3\textwidth
]{fig9b.eps}}
\hspace{0.5cm}
\subfigure{\includegraphics[width=0.3\textwidth
]{fig9c.eps}} 
\caption{
\label{fig9} (colour online)
Magnetic field, flux and current in 
the "tropics" ($z=360$) above the north antimonopole,
and in the equatorial plane ($z=0$). 
The north antimonopole is at $z=240$.}
\end{figure*}

In Figs. \ref{fig7} and \ref{fig8} we study the influence of temperature on the total field. 
As the temperature increases the flux tube between the two monopoles
narrows in the intermediate region and close to the critical temperature,
where the superconducting order vanishes, it approaches the normal phase regime where
close to the monopoles the field intensity extends away from the axis.
As the temperature increases the flux tubes narrow considerably until it breaks down
when the superconductor becomes unstable and the system becomes normal.
This happens both for $n=1$ and $n=0$. Again, the $n=0$ case simulates a situation
where the magnetic solenoids are being inserted from the border of the material such 
that it is favourable for the flux lines to emerge from the material.
As the monopoles get closer it is more favourable to establish a flux
tube between the monopoles.

We may consider different planes perpendicular to the string axis.
The flux contained in any
plane perpendicular to the strings axis is constant. In the regions far from the monopoles
the flux is almost entirely contained in the strings. This is shown in Fig. \ref{fig9}.
The flux profile is shown along a plane that intercepts the string and it is clear that
the flux reaches its asymptotic value very close to the string. In the region between the
monopoles the flux lines emerge and converge in the opposite charge monopole. In Fig. \ref{fig9}
it is clear that the flux grows smoothly from the origin, since in this case there is a 
considerable value of the magnetic field far from the axis. 
Also, we present results for $B_z$ and $J_{\varphi}$ also as a function of temperature.

In the central equatorial region ($z=0$) the magnetic field is maximum on the axis
and then decreases rapidly, on the scale of the penetration length. The field is compensated
by the supercurrents that also decay rapidly as we move away from the axis.

In a plane such that we are already on the strings, the field is finite and negative 
at the axis but increases to zero far from the string. There is a cancellation between
the field of the string and the field of the monopoles, as explained above.
The supercurrents vanish on the string, then go through a maximum at a finite distance
from the axis and then decrease far away. Note that the flux is already saturated
on the axis, then decreases and the asymptotic value is recovered far from the axis.

As the temperature increases the penetration length increases. As a consequence both
the increase and decrease of the field with distance sets on a larger scale as
the temperature increases. This is clearly shown on the various quantities plotted
in Fig. \ref{fig9}. 

\begin{figure}
\begin{picture}(350,315)(0,0)
\put(0,165){
\includegraphics[width=0.4\textwidth]{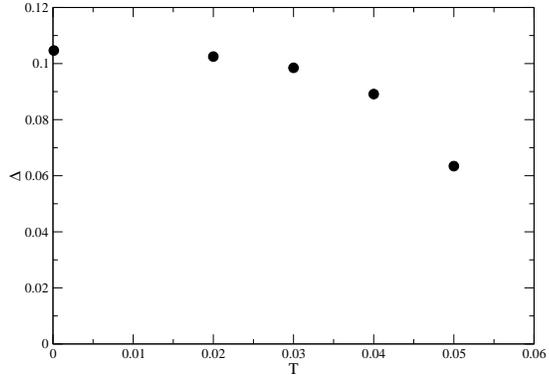}
}
\put(0,-5){
\includegraphics[width=0.4\textwidth]{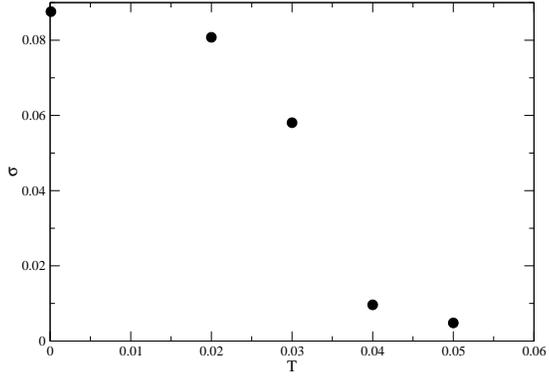}
}
\end{picture}
\caption{\label{fig10} 
a) Order parameter $\Delta$ as a function of temperature,
b) String tension $\sigma$ as a function of temperature.}
\end{figure}

Finally, we consider in Fig. \ref{fig10} the temperature dependence of the string
tension. Also, we show the temperature dependence of the bulk value of the
pairing function. We see that as the temperature increases both the pairing
function and the string tension decrease, particularly as the critical
temperature is approached. This critical temperature is defined as
the temperature when the pairing function vanishes.
Note that Fig. \ref{fig10} shows that the string tension is already quite small
when the pairing function is still finite, even though small and close to
the critical temperature.

\section{Conclusion}

Motivated by the superconductor model for the
confinement in QCD, we extended the Ball and Caticha
work to include both the microscopic fermionic degrees
of freedom of the superconductor and finite temperature.
We addressed two confining configurations, the $n=1$ 
configuration where the monopole is attracted by the antimonopole, 
and the $n=0$ configuration where the monopole is attracted 
by the boundary of the superconductor.
Our framework are the Bogoliubov-de Gennes equations
solved in the external field lines of a monopole-antimonopole
pair, including the associated semi-infinite magnetic strings.

Notice that this is not yet QCD, here we solve a condensed matter system.
The advantage of studying a solenoid-antisolenoid pair in
a superconductor is the more detailed nature of our results. For instance, 
we identify the well known localized fermionic states (unatainable via the Ginzburg-Landau
approach \cite{virtanen}), and we analyze the dependence
on the temperature of the string tension $\sigma$, of the fermion
pairing order parameter $\Delta$ 
and of the magnetic toplogical flux $\Phi$.

Nevertheless, this study is inspiring for the understanding
of confinement and of temperature and density effects in QCD 
\cite{DeGrand:1983fk,Teper:1985gi,DeTar:1987ar,Gottlieb:1987ac,
Bitar:1990si,Irback:1991eh,Koch:1994zt,Ejiri:1994uw,DeTar:1994jq,
Ichie:1994eg,Bernard:1996iz,Bernard:2001tz} 
or in Abelian QED
\cite{Chernodub:2001ws,Chernodub:2002gp}.
These effects are usually explored with Lattice QCD, Schwinger-Dyson
equations, or the Ginzburg-Landau equations.
We note that the present framework can also be extended to model QCD. 
Importantly, this extension
would be relevant to color superconductivity, where the quark 
density is large, and to the quark-gluon plasma, where
the temperature is finite.

Localized electron states exist in the region where the 
supercurrents provide a topological charge. As expected in
the Bogoliubov-de Gennes framework, we obtain one fermionic
localized state per angular momentum \cite{virtanen}. 

We find that the magnetic field in the vortex, say in the 
equatorial plane for $n=1$, decreases with increasing 
temperature. 
In what concerns the penetration length and the topological
magnetic flux, it is seen from the results that the penetration 
length increases with temperature (as is well known), while the 
topological magnetic flux decreases.
This decreases the string tension of the long 
distance monopole-antimonopole linear attraction.
Notice that the the string tension is already quite small
when the pairing function is still finite, even though small 
and close to the critical temperature.

\begin{thebibliography}{99}

\bibitem{Gross:1973id}
  D.~J.~Gross and F.~Wilczek,
  Phys.\ Rev.\ Lett.\  {\bf 30}, 1343 (1973).

\bibitem{Politzer:1973fx}
  H.~D.~Politzer,
  Phys.\ Rev.\ Lett.\  {\bf 30}, 1346 (1973).

\bibitem{Nielsen:1973cs}
  H.~B.~Nielsen and P.~Olesen,
  Nucl.\ Phys.\ B {\bf 61} (1973) 45.

\bibitem{Goto:1971ce}
  T.~Goto,
  Prog.\ Theor.\ Phys.\  {\bf 46} (1971) 1560.

\bibitem{Nambu:1974zg}
  Y.~Nambu,
  Phys.\ Rev.\ D {\bf 10}, 4262 (1974).

\bibitem{'tHooft:1974qc}
  G.~'t Hooft,
  Nucl.\ Phys.\ B {\bf 79}, 276 (1974).

\bibitem{Mandelstam:1974vf}
  S.~Mandelstam,
  Phys.\ Lett.\ B {\bf 53}, 476 (1975).

\bibitem{Dirac:1948um}
  P.~A.~M.~Dirac,
  Phys.\ Rev.\  {\bf 74} (1948) 817.


\bibitem{Baker:1984qh}
  M.~Baker, J.~S.~Ball and F.~Zachariasen,
  Phys.\ Rev.\ D {\bf 31}, 2575 (1985).

\bibitem{Baker:1991bc}
  M.~Baker, J.~S.~Ball and F.~Zachariasen,
  Phys.\ Rept.\  {\bf 209}, 73 (1991).

\bibitem{Wilson:1974sk}
  K.~G.~Wilson,
  Phys.\ Rev.\ D {\bf 10}, 2445 (1974).

\bibitem{Bali:1992ab}
  G.~S.~Bali and K.~Schilling,
  Phys.\ Rev.\ D {\bf 46}, 2636 (1992).

\bibitem{Bali:1994de}
  G.~S.~Bali, K.~Schilling and C.~Schlichter,
  Phys.\ Rev.\ D {\bf 51}, 5165 (1995)
  [arXiv:hep-lat/9409005].

\bibitem{Ivanenko:1990xu}
  T.~L.~Ivanenko, A.~V.~Pochinsky and M.~I.~Polikarpov,
  Phys.\ Lett.\ B {\bf 252}, 631 (1990).

\bibitem{Polikarpov:1987yr}
  M.~I.~Polikarpov and A.~I.~Veselov,
  Nucl.\ Phys.\ B {\bf 297}, 34 (1988).
  
\bibitem{Alford:1997zt}
  M.~G.~Alford, K.~Rajagopal and F.~Wilczek,
  %
  Phys.\ Lett.\ B {\bf 422}, 247 (1998)
  [arXiv:hep-ph/9711395].

\bibitem{DeRujula:1975ge}
  A.~De Rujula, H.~Georgi and S.~L.~Glashow,
  Phys.\ Rev.\ D {\bf 12}, 147 (1975).

\bibitem{Henriques:1976jd}
  A.~B.~Henriques, B.~H.~Kellett and R.~G.~Moorhouse,
  Phys.\ Lett.\ B {\bf 64}, 85 (1976).

\bibitem{Takahashi:2000te}
  T.~T.~Takahashi, H.~Matsufuru, Y.~Nemoto and H.~Suganuma,
  Phys.\ Rev.\ Lett.\  {\bf 86}, 18 (2001)
  [arXiv:hep-lat/0006005].

\bibitem{Takahashi:2002it}
  T.~T.~Takahashi and H.~Suganuma,
  Phys.\ Rev.\ Lett.\  {\bf 90}, 182001 (2003)
  [arXiv:hep-lat/0210024].

\bibitem{Okiharu:2004wy}
  F.~Okiharu, H.~Suganuma and T.~T.~Takahashi,
  Phys.\ Rev.\ Lett.\  {\bf 94}, 192001 (2005)
  [arXiv:hep-lat/0407001].

\bibitem{Okiharu:2004ve}
  F.~Okiharu, H.~Suganuma and T.~T.~Takahashi,
  Phys.\ Rev.\ D {\bf 72}, 014505 (2005)
  [arXiv:hep-lat/0412012].

\bibitem{Chodos:1974je}
  A.~Chodos, R.~L.~Jaffe, K.~Johnson, C.~B.~Thorn and V.~F.~Weisskopf,
  Phys.\ Rev.\ D {\bf 9}, 3471 (1974).

\bibitem{Jaffe:1975fd}
  R.~L.~Jaffe and K.~Johnson,
  Phys.\ Lett.\ B {\bf 60}, 201 (1976).

\bibitem{Goddard:1973qh}
  P.~Goddard, J.~Goldstone, C.~Rebbi and C.~B.~Thorn,
  Nucl.\ Phys.\ B {\bf 56} (1973) 109.

\bibitem{Bicudo:1989sh}
  P.~J.~d.~Bicudo and J.~E.~F.~Ribeiro,
  Phys.\ Rev.\ D {\bf 42}, 1611 (1990).

\bibitem{Ball:1987cf}
  J.~S.~Ball and A.~Caticha,
  Phys.\ Rev.\ D {\bf 37}, 524 (1988).

\bibitem{book} P.G. de Gennes, {\it Superconductivity of Metals and Alloys}
(Addison-Wesley, Reading, MA, 1989).


\bibitem{Gygi}
F.~Gygi and M. Schl\"uter,
Phys.\ Rev.\ B {\bf 43}, 7609 (1991)

\bibitem{vorticity} M. Cardoso, P. Bicudo and P.D. Sacramento, cond-mat/0509656.




\bibitem{virtanen} S.M.M. Virtanen and M.M. Salomaa, Phys.\ Rev.\ B {\bf 60}, 14581 (1999).



\bibitem{DeGrand:1983fk}
  T.~A.~DeGrand and C.~E.~DeTar,
  Nucl.\ Phys.\ B {\bf 225}, 590 (1983).

\bibitem{Teper:1985gi}
  M.~Teper,
  Phys.\ Lett.\ B {\bf 171}, 81 (1986).
  
\bibitem{DeTar:1987ar}
  C.~DeTar and J.~B.~Kogut,
  Phys.\ Rev.\ Lett.\  {\bf 59}, 399 (1987).
  
\bibitem{Gottlieb:1987ac}
  S.~A.~Gottlieb, W.~Liu, D.~Toussaint, R.~L.~Renken and R.~L.~Sugar,
  Phys.\ Rev.\ Lett.\  {\bf 59}, 2247 (1987).

\bibitem{Bitar:1990si}
  K.~M.~Bitar {\it et al.},
  Phys.\ Rev.\ D {\bf 43}, 2396 (1991).

\bibitem{Irback:1991eh}
  A.~Irback, P.~LaCock, D.~Miller, B.~Petersson and T.~Reisz,
  Nucl.\ Phys.\ B {\bf 363}, 34 (1991).

\bibitem{Koch:1994zt}
  V.~Koch,
  Phys.\ Rev.\ D {\bf 49}, 6063 (1994)
  [arXiv:hep-ph/9401284].

\bibitem{Ejiri:1994uw}
  S.~Ejiri, S.~i.~Kitahara, Y.~Matsubara and T.~Suzuki,
  Phys.\ Lett.\ B {\bf 343}, 304 (1995)
  [arXiv:hep-lat/9407022].
  
\bibitem{DeTar:1994jq}
  C.~DeTar,
  Nucl.\ Phys.\ Proc.\ Suppl.\  {\bf 42}, 73 (1995)
  [arXiv:hep-lat/9412010].
 
\bibitem{Ichie:1994eg}
  H.~Ichie, H.~Suganuma and H.~Toki,
  Phys.\ Rev.\ D {\bf 52}, 2944 (1995)
  [arXiv:hep-ph/9502278].
  
\bibitem{Bernard:1996iz}
  C.~W.~Bernard {\it et al.},
  Phys.\ Rev.\ Lett.\  {\bf 78}, 598 (1997)
  [arXiv:hep-lat/9611031].

\bibitem{Bernard:2001tz}
  C.~W.~Bernard {\it et al.},
  Phys.\ Rev.\ D {\bf 64}, 074509 (2001)
  [arXiv:hep-lat/0103012].
  
\bibitem{Chernodub:2001ws}
  M.~N.~Chernodub, E.~M.~Ilgenfritz and A.~Schiller,
  Phys.\ Rev.\ D {\bf 64}, 054507 (2001)
  [arXiv:hep-lat/0105021].

\bibitem{Chernodub:2002gp}
  M.~N.~Chernodub, E.~M.~Ilgenfritz and A.~Schiller,
  Phys.\ Rev.\ D {\bf 67}, 034502 (2003)
  [arXiv:hep-lat/0208013].
  



\end{thebibliography}
\end{document}